\documentstyle[11pt,epsfig]{article}
[1999/03/29 PSNFSS v.7.2
Times font as default roman: S Rahtz]

\begin{document}
\renewcommand{\thefootnote}{\fnsymbol{footnote}}
\sloppy
\newcommand{\rp}{\right)}
\newcommand{\lp}{\left(}
\newcommand \be  {\begin{equation}}
\newcommand \bea {\begin{eqnarray}}
\newcommand \ee  {\end{equation}}
\newcommand \eea {\end{eqnarray}}

\title{Comment on recent claims by Sornette and Zhou}
\author{Anders Johansen 
\\ (1) Ris\o \ National Laboratory, Department of Wind Energy \\
Frederiksborgvej 399, P.O. 49, DK-4000 Roskilde, Denmark \\
e-mail: anders.johansen@risoe.dk,  URL: http://www.risoe.dk/vea/staff/andj/ }

\maketitle

Owing to a large number of press releases in which my work has been
heavily cited in support of the recent SP500 prediction by Sornette and 
Zhou (SZ), I feel it necessary to comment on this work and their follow-up 
preprint \cite{ZS}. 
The predictions by SZ regarding the future behaviour of in particular the SP500
has received quite some attention and a substantial part of the evidence 
presented supporting the predictions of SZ is based on my numerical analysis
of the Nikkei in the period 1990-2000\cite{Nikkei}. Hence, I feel urged to 
present my own view on the log-periodic power law (LPPL) analysis of the 
financial markets (FM) made by SZ and in particular on the claims of LPPL 
behaviour in the FM in general and the SP500 in particular as well as the 
predictions that SZ derive from their analysis. In 1996 and 1997 two groups 
independently proposed that power laws with complex exponents, {\it e.g.},
\bea \label{lpeq}
p(t) \approx A + B (t_c-t)^z 
+C(t_c-t)^z\cos\lp\omega\log(t_c-t)+\phi\rp
\eea
were relevant modeling tools for the description of price $p(t)$ increases 
a few years prior to very large crashes \cite{SJB,FF}. The background for
the original suggestion of LPPL signatures in the financial markets was an 
analogy between second order phase transitions and rupture, in this context a 
``rupture in market belief''. Furthermore, it was proposed that the domain of 
the power law exponent should not be restricted to real values only. 
Consequently, the analogy was not confined to a pure power law behaviour but 
allowed for a power law behaviour decorated by so-called log-periodic 
oscillations, retrospectively to be seen as a quite provocative claim 
\cite{Lalouxrep}. Disregarding the rupture analogy (for which the empirical 
evidence is scarce), one may also consider the proposed frame work simply as 
an {\it Ansatz}
\bea \label{ansatz}
\frac{d F\lp x\rp}{d \log x} = \alpha F\lp t\rp + \mbox{higher order terms}
\eea
for the dynamical rescaling of a price (or some related quantity) $F\lp x\rp$
as a function of ``time to the crash'' $x = t_c -t$. Such an {\it Ansatz} 
approach is not uncommon in the field of critical phenomena. Before commenting 
further on the recent claims by SZ, I should stress that the 
present author with D. Sornette in \cite{epsilon} has presented a synthesis of
two independent research directions, namely that of LPPL analysis on the one 
hand and the ``outlier'' classification of the largest negative market events 
on the other. In essence, that paper propose an objective criterion for the 
selection of events which {\it could} have LPPL precursors. The conclusion of 
that analysis is that a large negative market event which classifies and an 
outlier is either preceded by an LPPL speculative bubble {\it or} an 
unsuspected (to judge from the market response) historical event. (This
does not exclude the possibility of ``other precursory events.'') The 
statistical evidence for this proposition is quite convincing. Furthermore, a 
statistical analysis of what has been referred to as the two  ``physical 
variables'' $z$ and $\omega$ \cite{bali,epsilon} has been presented. 
(The background for the term  ``physical variables'' is that the variables 
$A,B,C,\phi$ are nothing but units and $t_c$ is event specific.) In this 
context, it is worth noting that the ``double cosine'' equation
proposed by 
SZ, {\it i.e.}, 
\bea \label{2cosine}
p(t) \approx A + B (t_c-t)^z  +C(t_c-t)^z\cos\lp\omega\log(t_c-t)+\phi_1 \rp 
+ D(t_c-t)^z\cos\lp 2 \omega\log(t_c-t)+\phi_2 \rp
\eea
has different phases. Since the phases in eq.'s (\ref{lpeq}) and 
(\ref{2cosine}) simply are time units (changing the time units of the 
data from for example days to months only changes radically the value of
$\phi$, as it should, and not the other variables) needed because of the 
$\log(t_c-t) -\phi = \log\lp (t_c-t)/\phi'\rp$, a sound theoretical 
justification for such a ``phase-shift'' between the ``first and the second 
harmonics'' (to use the terminology of SZ) is lacking to say the least. I also 
wish to stress that the conclusion of the analysis of \cite{bali} of the
values obtained for the physical variables $z$ and $\omega$ (based on a 
Gaussian null-hypothesis for the pdf) including over 30 case studies is that 
\bea \label{restric}
\omega \approx 6.36 \pm 1.56 \hspace{10mm} z \approx 0.33 \pm 0.18.
\eea
Unfortunately, a comparison between this statistical estimate and the more
recent analysis presented by SZ is completely absent. In fact, making a similar
statistical analysis of the results presented in \cite{ZS} on
anti-bubbles (since SZ advocates the existence of bubbles and
anti-bubbles from a symmetry perspective, a compassion between the
estimates \protect\ref{restric} for bubbles and their results for anti-bubbles
is necessary and easy) yields a uniform distribution of the physical variable 
$\omega$ with high probability. What I find quite peculiar is that I with 
Sornette proposed in \cite{manifesto} a set of very basic assumptions which a 
LPPL analysis of financial data should full-fill: 1) Landau expansions, 
{\it i.e.,}, eq. (\ref{ansatz}); 2) Bounded rationality (or ``conservation 
laws''), {\it e.g.}, prices should not go to infinity as they do in the the 
so-called bullish anti-bubble of SZ, where they accept $B>0$; 3) Symmetry 
considerations, {\it e.q.}, Statistical long-term asymmetry where market drops 
are fast and market increases are slow; 4) Probabilistic framework, due to the 
fact that the financial markets are a non-closed system, which however may 
behave as a semi-closed system over time; 5) Most importantly, {\it any 
validation} of a model must come from the data, {\it e.g.}, a statistical 
analysis of the empirical results obtained from the numerous case studies 
presented in the literature. My main objection to the work of SZ (as well as 
those of others others) is that the fundamental concept of criticality has
apparently been abandoned, {\it e.g.}, many case studies have been presented by
 DZ (among others) where $z\approx 1$ seems to have become so natural that 
nobody seems to question it anymore. Another violation of 
the framework proposed above is that SZ now have changed the control parameter 
$x=t_c -t$ (or $-x$ for anti-bubbles) in eq. (\ref{lpeq}) to  $x=|t_c -t|$. 
This means that another restriction coming form the data has
disappeared.  
A comment 
on the so-called ``fractal'' concept, (LPPL within LPPL) where authors have 
claim such a signature on a single case study in which the analysis by eye has 
identified an single example. As I previously performed an extensive analysis 
of such ``fractal structures'' mainly in collaboration with Matt Lee, another 
former post doc of Sornette. We analyzed over 10 different statistical 
indexes of stock, currencies and bonds without any conclusive results. Each 
data set had a length of 2-4 years. We did get a slightly (1-5\%) better binary
(``up or down'') prediction rate for the US market, the DAX and the FTSE on a 
two to four week prediction horizon. As described in detail in \cite{SJ2000} 
the real success however was with a LPPL analysis on time scales of 1-2 years
using the same time period for the data. 
It should be stressed that one of the crucial criteria for this success rate 
of crash and LPPL bubble identification was the restriction $B<0$ as well as 
the bound on the physical variables $z$ and $\omega$ corresponding 
(\ref{restric}). Most importantly, I wish to stress that the postulated 
similarity between the behaviour of the Nikkei index in the period 1990-2000 
years with that of the SP500 in the past couple of years is completely 
unsubstantiated in the papers by SZ. I find it inappropriate that my numerical 
analysis presented in \cite{Nikkei} can be used to support the present 
prediction of SZ. Just to mention three serious discrepancies between the two 
countries (Japan and the U.S.A.), the value of the log-periodic frequency 
differs by a factor of 2 despite the ``double cosine'' eq.(\ref{2cosine}.  
Furthermore, the Nikkei did not go through a "classical" LPPL bubble prior 
to the onset of the anti-bubble as the U.S. market (Nasdaq) did. (A real-estate
bubble seems to be the favorite explanation for this. The statistical evidence 
so far on anti-bubbles seems that external shocks such as, {\it e.g.},the
effect of the the burst of the Asian bubble of '97 on the major western stock
markets, are ``the cause'' and not internally generated.) Last, but not least,
the Nikkei analysis was based on 9  years of data, with the first data point
objectively being chosen as the peak of the market price. The present
SP500 prediction of SZ  is not consistent with these facts.

 All papers by the author can be retrieved from
http://www.risoe.dk/vea/staff/andj/pub.html.


\begin{thebibliography}{}

\bibitem{ZS} D. Sornette and W. Zhou, Quantitative Finance 2 (6), 468-481 
(2002); Evidence of a Worldwide Stock Market Log-Periodic Anti-Bubble Since 
Mid-2000, cond-mat/0212010; Renormalization Group Analysis of the 2000-2002 
anti-bubble in the US SP 500 index, physics/0301023
\bibitem{manifesto} A. Johansen and D. Sornette, Eur. Phys. J. B9, 
pp. 167-174 (1999). 
\bibitem{SJB} D. Sornette, A. Johansen and J.P. Bouchaud,  
J. Phys. I. France 6 pp. 167-175 (1996)

\bibitem{bali} A. Johansen  Characterization of large price variations in 
financial markets, To be published in Physica A.

\bibitem{FF} J.A. Feigenbaum and P.G.O. Freund, (1998), Modern Physics Letters 
B 12: 57. J. A. Feigenbaum and P.G.O. Freund, (1996), Int. J. Moder Phys. 
B 10: 3737

\bibitem{SJ2000} D. Sornette and A. Johansen, Quantitative Finance 
vol.1 pp. 452-471 (2001)

\bibitem{epsilon} A. Johansen and D. Sornette, Endogenous versus 
Exogenous Crashes in Financial Markets. Submitted to Journal of Economic 
Dynamics and Control.

\bibitem{Lalouxrep} A. Johansen, Europhys. Lett. 60 (5), pp.809-810
(2002) and references therein.


\bibitem{Nikkei}A. Johansen and D. Sornette, Int. J. Mod. Phys. 10, pp.
563 -575 (1999),  A. Johansen and D. Sornette,  Int. J. Mod. Phys. 
C 11 no. 2 pp. 359-364 (2000)  

\end{thebibliography}
\end{document}